\documentclass[aps,pre,twocolumn,showpacs,groupedaddress]{revtex4}
\usepackage{graphicx}
\usepackage{dcolumn}
\usepackage{bm}
\usepackage{amssymb}
\usepackage{amsmath}
\usepackage{color} 

\begin{document}

%\preprint{}

%Title of paper
\title{Non-uniqueness of local stress of three-body potentials in molecular simulations} 

\author{Koh M. Nakagawa and Hiroshi Noguchi}
\email[]{noguchi@issp.u-tokyo.ac.jp}
\affiliation{
Institute for Solid State Physics, University of Tokyo,
 Kashiwa, Chiba 277-8581, Japan}

\date{\today}

\begin{abstract}
Microscopic stress fields are widely used in molecular simulations to understand mechanical behavior.
Recently, decomposition methods of multibody forces to central force pairs between the interacting particles
have been proposed. Here, we introduce a force center of a three-body potential
and propose different force decompositions that 
also satisfy the conservation of translational and angular momentum.
We compare the  force decompositions by stress-distribution magnitude and
discuss their difference in the stress profile of a bilayer membrane  
using coarse-grained and atomistic molecular dynamics simulations.
\end{abstract}

\pacs{87.10.Tf,83.10.Rs,87.16.D-}
%% 87.	Biological and medical physics
%% 87.10.-e	General theory and mathematical aspects
%% 87.10.Tf	Molecular dynamics simulation
%% 83.	Rheology
%% 83.10.-y	Fundamentals and theoretical
%% 83.10.Rs 	Computer simulation of molecular and particle dynamics
%% 87.16.-b	Subcellular structure and processes
%% 87.16.D-	Membranes, bilayers, and vesicles

%\keywords{}

\maketitle

\section{Introduction}

%% Local pressure profile calculation long history
%%% for two body potential
The stress tensor is a fundamental quantity that connects discrete molecular systems and continuum mechanics.
The calculation of the local stress field from molecular simulations has
a long history~\cite{ irving1950statistical, noll1955herleitung,hardy1982formulas,schofield1982statistical, todd1995pressure, goetz1998computer,heinz2005calculation,admal2010unified,admal2014results,admal2011stress, admal2016non,vanegas2014importance,torres2015examining,torres2016geometric}.
Irving and Kirkwood introduced the microscopic stress tensor formula based on non-equilibrium statistical mechanics~\cite{irving1950statistical}, 
following which a rigorous mathematical formula was proposed by Noll~\cite{noll1955herleitung}.
In the following, we refer to their procedure as the Irving-Kirkwood-Noll (IKN) procedure, as stated by Admal and Tadmor~\cite{admal2010unified}.
Hardy introduced the spatial averaging of the stress tensor using weighting functions
to improve statistics~\cite{hardy1982formulas}.
However, these procedures are limited to systems in which interactions consist of pairwise forces. 

The method to map the stress of multibody potentials into the continuum space has been debated.
Multibody potentials have been frequently used in molecular simulations.
Bending and dihedral potentials, which are widely used, are three- and four-body potentials, respectively. 
The interaction between adjacent dihedrals is represented by five-body correction map (CMAP) potential in the CHARMM force field~\cite{mackerell2004extending,torres2015examining}. 
A curvature potential in meshless membranes is a function of three rotational invariants of the weighted gyration tensor and 
produces $n$-body forces, 
where $n$ depends on the local density~\cite{nogu06}.
Since most of the multibody forces are not central forces between particles, 
the IKN procedure cannot be directly applied to them. 
Note that multibody hydrophobic potentials as a function of the local hydrophobic particle density for proteins~\cite{taka99} 
and membranes~\cite{nogu06,nogu01a}
give central forces between particles so that their stress can be calculated directly using the IKN procedure.

Goetz and Lipowsky proposed a decomposition procedure 
for multibody potentials~\cite{goetz1998computer} based on Schofield and Henderson's procedure~\cite{schofield1982statistical}. 
Multibody forces are decomposed into
pairwise (non-central) forces, and the IKN procedure is applied to each
decomposed force pair.
We refer to this method as Goetz-Lipowsky decomposition (GLD).
However, GLD does not satisfy the strong law of action and reaction, as pointed out by Admal and Tadmor~\cite{admal2010unified,admal2014results},
while it satisfies the weak law of action and reaction:
GLD conserves translational momentum but not angular momentum. 
Consequently, the stress tensor is not symmetric. To overcome this problem, central force
decomposition (CFD) was proposed~\cite{admal2010unified, admal2014results, admal2016non, torres2015examining, vanegas2014importance}. 
The forces decomposed using CFD satisfy the strong law of action and reaction so that
the stress tensor is symmetric by construction.
The original CFD is limited to three- or four-body forces because there
exists a unique force decomposition for only up to four-body forces.
For $n$-body forces with $n\ge 5$,
the number of degrees of freedom $3n-6$ is less than the number of pairs $n(n-1)/2$ in the three-dimensional (3D) space.
Very recently, the generalization to more than four-body forces, which is called a covariant CFD (cCFD), 
was introduced by Torres-S{\'a}nchez {\it et al.}~\cite{torres2015examining,torres2016geometric}.
The application to a structural coiled-coil protein with the five-body CMAP potential was demonstrated~\cite{torres2015examining}.

In this paper, we discuss non-uniqueness in the force decomposition of three-body forces in classical mechanics.
Three-body forces can be uniquely decomposed by CFD.
However, we will show different decompositions, which also satisfy the strong law of action and reaction.
A force center can be uniquely defined for three-body forces,
and the forces are decomposed into central force pairs between interacting particles and the force center.
To combine this decomposition and CFD, the position of the force center can be arbitrarily taken. 
This non-uniqueness is related with the non-unique potential-energy extension discussed in Ref.~\cite{admal2010unified,admal2014results,admal2011stress}.
It is a specific case of the degeneracy of four-body forces into 2D space.
We will discuss the choice of this center position by the stress distribution.
Although two-body forces can also similarly be decomposed,
the IKN procedure always gives the minimum stress distribution.
In contrast, the stress distribution of three-body forces depends on the type of the forces.
We will also discuss the influence of the resolution of simulation models.

For an application of the force decomposition, we investigated a bilayer
membrane using coarse-grained and atomistic molecular dynamics (MD) simulations.
The stress profile along the normal direction has been widely calculated
in the molecular simulations of lipid membranes.
Two opposing forces, interfacial tension and steric repulsion, produce the inhomogeneous stress inside the bilayers~\cite{israelachvili2011intermolecular}.
This inhomogeneity is a key property of the bilayers because it determines the area per lipid molecule~\cite{israelachvili2011intermolecular},
spontaneous
curvature~\cite{rozycki2015spontaneous,venable2015mechanical,orsi2008quantitative,orsi2011elba,hu2013gaussian},
Gaussian curvature
modulus~\cite{safran2003statistical,helfrich1994lyotropic,venable2015mechanical,hu2013gaussian,hu2012determining,orsi2008quantitative,orsi2011elba,shinoda2011free},
and function of the mechanosensitive channel~\cite{sotomayor2004molecular,vanegas2014force}. 
Since the stress profile cannot be obtained experimentally~\cite{fravnova2014can,templer1999sensing},
estimation using molecular simulations is important.
Recently, however, Torres-S{\'a}nchez {\it et al.}
reported that the stress profile is strongly dependent on the force decomposition method~\cite{torres2015examining}.
The dihedral forces give the largest contribution to the stress profile by CFD.
We will show that the stress profile is largely dependent on the decomposition of bending forces.

In Sec.~\ref{sec:dec}, we discuss the force decomposition method.
After introducing the existing decomposition method,
we describe the alternative decomposition method for three-body forces.
As an example, we show the decomposition for an area potential
and a bending potential.
The area potential is one of the simplest three-body potentials
and is connected to continuum mechanics in a straightforward manner.
The bending potential is the most widely used three-body potential.
In Sec.~\ref{sec:mem}, the bilayer membrane is examined.
The  stress profile and Gaussian curvature modulus are calculated for different decomposition methods.
The  discussion and summary are given in Sec.~\ref{sec:dis} and \ref{sec:sum}, respectively.

\section{Force Decomposition}\label{sec:dec}

\subsection{Irving-Kirkwood-Noll procedure}\label{sec:ikn}

Stress averaged over the entire simulation box is given by the virial as
\begin{eqnarray}
{\boldsymbol \sigma} &=& {\boldsymbol \sigma}_{\rm K} + {\boldsymbol \sigma}_{\rm U} ,\\ \label{eq:stK}
{\boldsymbol \sigma}_{\rm K} &=& - \frac{1}{V}\sum_i \langle m_i {\bf v}_i \otimes {\bf v}_i \rangle , \\
{\boldsymbol \sigma}_{\rm U} &=&  - \frac{1}{V}\sum_i  \langle  {\bf f}_i \otimes {\bf r}_i \rangle, \label{eq:stU} \\ \label{eq:stUm}
&=&  \frac{1}{V}\sum_{n=2}^{N} \sum_{k_n=1}\sum_{i=1}^{n}  \langle \frac{\partial U_{k_n}}{\partial {\bf r}_{k_n,i}} \otimes ({\bf r}_i - {\bf r}_{k_n, 0}) \rangle,
\end{eqnarray}
where $m_i$, ${\bf r}_i$, and ${\bf v}_i$ are the mass, position, and velocity of the $i$-th particle
and ${\bf f}_i= - \partial U/\partial {\bf r}_i$.
The symbol $\otimes$ denotes a tensor product
and $\langle ... \rangle$ denotes a statistical average.
This global stress is uniquely determined even for multibody forces.
The potential contribution ${\boldsymbol \sigma}_{\rm U}$ can be rewritten with Eq.~(\ref{eq:stUm})
using cluster expansion~\cite{torres2016geometric} as
\begin{equation}
U({\bf r}_1,...,{\bf r}_N)= \sum_{n=2}^{N} \sum_{k_n=1} U_{k_n}({\bf r}_{k_n,1},...,{\bf r}_{k_n,n}),
\end{equation}
where each $U_{k_n}$ is an $n$-body potential that is invariant under translation and rotation.
The origin ${\bf r}_{k_n, 0}$ of the positions can be taken differently for each $U_{k_n}$,
as expressed in Eq.~(\ref{eq:stUm}).
Each origin can be arbitrarily chosen but a position close to interacting particles is preferred to reduce numerical errors,
particularly for large-scale simulations.
When the origin is set to the position of one of the interacting particles,
the potential stress of the pairwise potentials takes the well-known form,
\begin{equation}
    {\boldsymbol \sigma}_{\rm U, pair} = -\dfrac{1}{V} \sum_{i<j} \langle f_{ij}  \hat{\bf r}_{ij} \otimes {\bf r}_{ij}\rangle,
  \end{equation}
where $f_{ij}= - \partial U_{\rm {k_2}}/\partial r_{ij}$,  ${\bf r}_{ij}={\bf r}_{i}-{\bf r}_{j}$, 
 $r_{ij}= |{\bf r}_{ij}|$, and $\hat{\bf r}_{ij} = {\bf r}_{ij}/r_{ij}$.
Under a periodic boundary condition, 
the periodic image is used instead of the original position
when the potential interaction crosses the periodic boundary.

For pairwise interactions,
the local stress at a position ${\bf x}$ is given by the IKN procedure
as~\cite{irving1950statistical,noll1955herleitung}
\begin{eqnarray}
{\boldsymbol \sigma}({\bf x}) &=& {\boldsymbol \sigma}_{\rm K}({\bf x})
                                  +{\boldsymbol \sigma}_{\rm U}({\bf x})
                                  ,\\ \label{eq:stKx}
{\boldsymbol \sigma}_{\rm K}({\bf x}) &=& - \sum_i \langle m_i {\bf v}_i \otimes
                                          {\bf v}_i \delta({\bf r}_i - {\bf x})
                                          \rangle , \\ \label{eq:stUx}
{\boldsymbol \sigma}_{\rm U}({\bf x}) &=& - \sum_{i< j} \langle f_{ij}  \hat{\bf
                                          r}_{ij}  \otimes {\bf r}_{ij} B({\bf
                                          r}_i, {\bf r}_j, {\bf x})\rangle,
\end{eqnarray}
where $B({\bf r}_i, {\bf r}_j, {\bf x})= \int_0^1 \delta[(1-s){\bf r}_i+s{\bf r}_j- {\bf x}]ds$.
The force propagates along the line segment between ${\bf r}_{i}$ and ${\bf r}_{j}$.
This local stress tensor is symmetric: $\sigma_{\alpha\beta}({\bf x})=\sigma_{\beta\alpha}({\bf x})$ for $\alpha,\beta \in \{x, y, z\}$.

\subsection{Central Force and Geometric-Center Decompositions}\label{sec:prof}

When the multibody force is decomposed into pairwise forces between interacting particles,
the IKN procedure for pairwise forces is applicable.
Therefore, decomposition methods to pairwise forces have been focused upon.
Goetz and Lipowsky proposed  a decomposition (GLD), ${\bf f}_{ij}=({\bf f}_{i} - {\bf f}_{j})/n$, 
for $n$-body forces~\cite{goetz1998computer}.
This decomposition conserves the translational momentum but does not conserve the angular momentum,
since the force ${\bf f}_{ij}$ is not generally parallel to ${\bf r}_{ij}$.

\begin{figure}
 \centering
 \includegraphics[width=80mm]{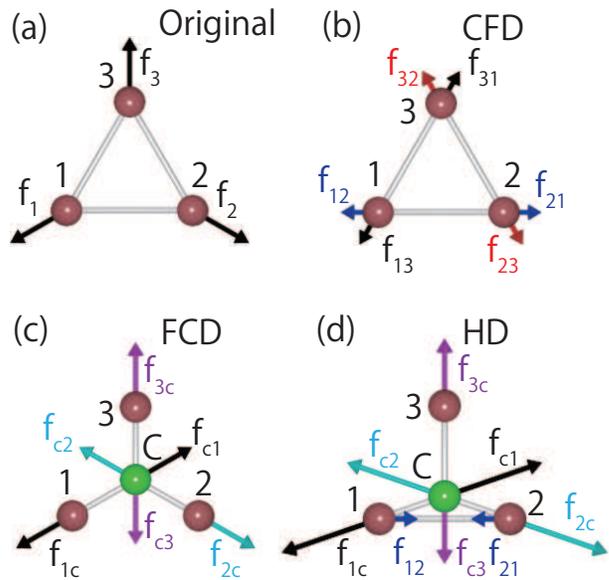}
 \caption{
   (Color online)
   Force decomposition for area expansion forces.
   (a) Original forces. (b) CFD. (c) FCD. (d) HD.
   The light gray (green) sphere represents the force center, ${\bf r}_c$.
}
 \label{fig:ar_force}
\end{figure}

\begin{figure}
 \centering
 \includegraphics[width=80mm]{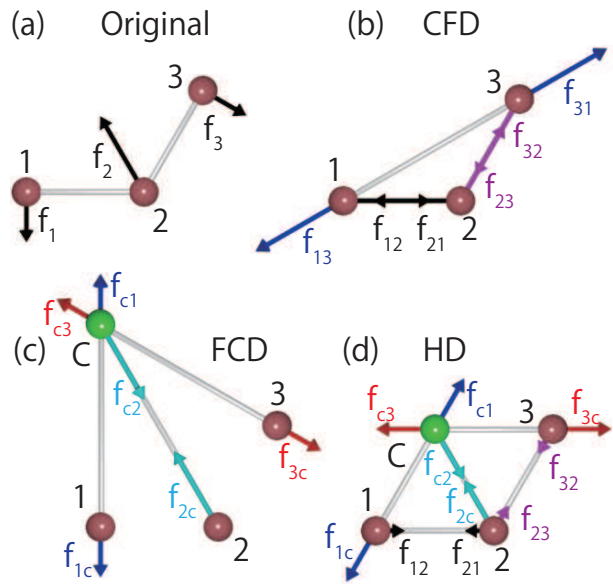}
 \caption{
   (Color online)
   Force decomposition for bending forces on $\theta_{123}$.
   (a) Original forces. (b) CFD. (c) FCD. (d) HD.
   The same color notation as Fig.~\ref{fig:ar_force} is employed.
}
 \label{fig:bend_force}
\end{figure}

In order to satisfy the conservation of the angular momentum as well,
Admal and Tadmor proposed the decomposition to central forces between  interacting particles (CFD)~\cite{admal2010unified,admal2014results}.
Three-body forces can be uniquely decomposed by CFD:
\begin{eqnarray}
{\bf f}_1 &=&  f_{12}\hat{\bf r}_{12} + f_{13}\hat{\bf r}_{13} , \nonumber \\ \label{eq:f123}
{\bf f}_2 &=&  f_{23}\hat{\bf r}_{23} + f_{12}\hat{\bf r}_{21} , \\
{\bf f}_3 &=&  f_{13}\hat{\bf r}_{31} + f_{23}\hat{\bf r}_{32} . \nonumber
\end{eqnarray}
Since the translational and angular momenta are conserved, 
${\bf f}_1+ {\bf f}_2+ {\bf f}_3={\bf 0}$  and ${\bf f}_1\times {\bf r}_1 + {\bf f}_2\times {\bf r}_2+ {\bf f}_3\times {\bf r}_3={\bf 0}$.
For $f_{12}>0$, $f_{12}$ is a repulsive force between ${\bf r}_1$ and ${\bf r}_2$.
From Eq.~(\ref{eq:f123}), $f_{12}$ is given by 
\begin{equation}\label{eq:f12a}
f_{12} = \frac{{\bf f}_1\cdot \hat{\bf r}_{12} - ({\bf f}_1\cdot \hat{\bf r}_{13})(\hat{\bf r}_{12}\cdot \hat{\bf r}_{13})}{1-(\hat{\bf r}_{12}\cdot \hat{\bf r}_{13})^2} ,
\end{equation}
or
\begin{equation}\label{eq:f12a}
f_{12} = \frac{1}{2}\Big( \frac{{\bf f}_1\cdot (\hat{\bf r}_{12}+ \hat{\bf r}_{13})}{1+\hat{\bf r}_{12}\cdot \hat{\bf r}_{13}  } + \frac{{\bf f}_2\cdot (\hat{\bf r}_{23}+ \hat{\bf r}_{21})}{1+\hat{\bf r}_{21}\cdot \hat{\bf r}_{23}  } - \frac{{\bf f}_3\cdot (\hat{\bf r}_{31}+ \hat{\bf r}_{32})}{1+\hat{\bf r}_{31}\cdot \hat{\bf r}_{32}  } \Big). 
\end{equation}
Similarly, $f_{13}$ and $f_{23}$ are given.
Equation~(\ref{eq:f12a}) is recommended for numerical calculations, 
since it gives smaller numerical errors 
when two angles of $\triangle 123$ are close to null and the third is close to $\pi$. 
Alternatively, these force pairs can be derived directly 
from $f_{12}=-\partial U_{k_3}/\partial r_{12}|r_{13},r_{23}$ \cite{admal2010unified}
as demonstrated for the area and bending potentials in 
Appendix~\ref{sec:area} and \ref{sec:bend}, respectively.
The CFDs of the area expansion and bending forces are 
shown in Figs.~\ref{fig:ar_force}(b) and \ref{fig:bend_force}(b), respectively.
The three interacting particles form a triangle and lie on a plane
so that the forces ${\bf f}_1$, ${\bf f}_2$, and ${\bf f}_3$ are along this plane
owing to the conservation of translational and angular momenta. 
Hence, we can consider the 2D space without loss of generality.

Alternatively, 
Heinz {\it et al.} proposed a decomposition method that uses the geometric center, $\sum_i^{n'} {\bf r}_i/n'$, of 
$n'$ interacting particles in a divided cell for an $n$-body potential ($n'<n$)~\cite{heinz2005calculation}.
In this decomposition, the angular momentum is not conserved.
The geometric center is determined only by the positions and has no relation to the force balance.
Hence, the geometric center can significantly deviate from the positions where the forces act.
For example, 
when great forces act only on two particles in $n$-body forces, {\it i.e.}, $|{\bf f}_{i}| \gg |{\bf f}_{j}|$ ($i=1$, $2$, and $j\ge 3$),
the resultant stress should be close to that of the pairwise forces between ${\bf r}_1$ and ${\bf r}_2$.
However, the geometric center can be far from the line segment between ${\bf r}_1$ and ${\bf r}_2$.
Thus, a center position should be determined by the force balance, 
or a specific force decomposition should be employed for a chosen center position to satisfy the force balance.
We consider the center position with the decomposition to satisfy the strong law of the action and reaction in Sec.~\ref{sec:fcd}.

One may consider the center of mass as an alternative candidate for the
center position.
However, the potential stress term ${\boldsymbol \sigma}_{U}$ is not dependent on mass distribution in thermal equilibrium.
One can calculate ${\boldsymbol \sigma}_{U}$ using a Monte Carlo
simulation, in which the mass distribution is not required at all.
Since the values of the particle masses are arbitrary but positive,
the center of mass lies inside the convex polyhedron (triangle for three-body forces) formed by interacting particles.
As described below, it is important whether the center position for
force decomposition is inside or outside the triangle for three-body forces.

\subsection{Force Center and Hybrid Decompositions}\label{sec:fcd}

We consider the alternative decompositions of three-body forces.
As mentioned above, the forces are uniquely determined by CFD for three-body forces.
However, when one more position is taken into account, the forces are not uniquely determined.
For three-body forces,
three lines drawn along the force vectors ${\bf f}_i$ from the particle positions ${\bf r}_i$ ($i \in {1,2,3}$)
always meet at one position owing to the angular-momentum conservation.
We refer to this position as the force center, ${\bf r}_c$.
It is determined as
\begin{eqnarray} \label{eq:fc}
{\bf r}_c &=& \frac{1}{q}\big( \tilde{f}_{12}\tilde{f}_{13}{\bf r}_1+\tilde{f}_{12}\tilde{f}_{23}{\bf r}_2+\tilde{f}_{13}\tilde{f}_{23}{\bf r}_3 \big)\\
q &=& \tilde{f}_{12}\tilde{f}_{13} +\tilde{f}_{12}\tilde{f}_{23}+\tilde{f}_{13}\tilde{f}_{23}
\end{eqnarray}
where $\tilde{f}_{ij}={f}_{ij}/r_{ij}$ and ${f}_{ij}$ are the forces obtained by CFD.
The sign of the denominator $q$ determines the region of the force center as described later.
Using the force center, the forces are decomposed into three central force pairs $f_{ic}\hat{\bf r}_{ic}={\bf f}_i$ 
between ${\bf r}_{i}$ and ${\bf r}_c$ for $i \in \{1,2,3\}$ [see Figs.~\ref{fig:ar_force}(c) and \ref{fig:bend_force}(c)].
We refer to this decomposition as force-center decomposition (FCD).
Since these are central forces, the strong law of action and reaction is satisfied
and the symmetric local stress tensor is obtained by the IKN procedure for these decomposed forces.

\begin{figure}
 \centering
 \includegraphics[width=55mm]{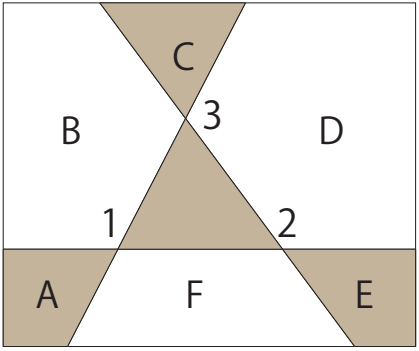}
 \caption{
 (Color online)
 Six exterior regions (A--F) of the triangle $\triangle 123$. 
Shaded and white regions correspond to $q > 0$ and $q < 0$, respectively.
 }
 \label{fig:q}
\end{figure}

When three force pairs have the same sign ($f_{12}>0$, $f_{13}>0$, $f_{23}>0$  or $f_{12}<0$, $f_{13}<0$, $f_{23}<0$),
the  force center lies in the interior region of the triangle 
$\triangle 123$ and $q>0$.
For an expansion force as shown in Fig.~\ref{fig:ar_force},
the decomposed forces in CFD and FCD can be physically interpreted as
line (surface) tension on the edge of the triangular region
and  pressure of the interior region on the particles, respectively.

The exterior region can be divided into six regions as shown in Fig.~\ref{fig:q}. 
When $f_{12}f_{13}>0$ and $f_{12}f_{23}<0$,
 the force center lies in the region A or D for $q>0$ or $q<0$, respectively.
For bending potentials as a function of the angle $\theta_{123}=\cos^{-1}(\hat{\bf r}_{12}\cdot\hat{\bf r}_{32})$,
${\bf r}_{c}$ always lies outside the triangle and $q<0$.
As $\theta_{123}$ becomes closer to $\pi$,  
 ${\bf f}_1$ and  ${\bf f}_3$ approach parallel lines
so that ${\bf r}_{c}$ becomes further from the particle positions.
The details of decomposition for the area and bending potentials are described in
Appendix~\ref{sec:area} and \ref{sec:bend}, respectively.

The force center position can be moved by combining FCD with CFD.
We refer to this combined decomposition as hybrid decomposition (HD).
If necessary to distinguish them, the force center in FCD is called the
original force center ${\bf r}_{c0}$.
In HD, the force pair of each edge of $\triangle 123$ is divided into FCD and CFD components as $f_{12}^{\rm all}=f_{12}^{\rm FC}+f_{12}^{\rm CF}$.
The force center ${\bf r}_c$ is determined by Eq.~(\ref{eq:fc}) with the FCD components $f_{12}^{\rm FC},f_{13}^{\rm FC}$, and $f_{23}^{\rm FC}$.
For example, Fig.~\ref{fig:ar_force}(d)
shows the decomposition into three force pairs with ${\bf r}_c$ and one force pair along ${\bf r}_{12}$.
When the contribution of force $f_{12}$ to FCD increases (decreases),
the hybrid force center ${\bf r}_{c}$ is further (closer) to ${\bf r}_3$  than the original force center ${\bf r}_{c0}$ [see Eq.~(\ref{eq:fc})].
Fig.~\ref{fig:bend_force}(d) shows HD combining FCD with two force pairs along ${\bf r}_{12}$ and ${\bf r}_{23}$.
If the force center lies on the edge of the triangle $\triangle 123$, the resultant decomposition coincides with CFD
(if ${\bf r}_{c}$ lies in the middle of the line segment between ${\bf r}_{2}$
and ${\bf r}_{3}$, then $f_{1c}=0$).

\begin{figure}
  \centering
  \includegraphics[width=90mm]{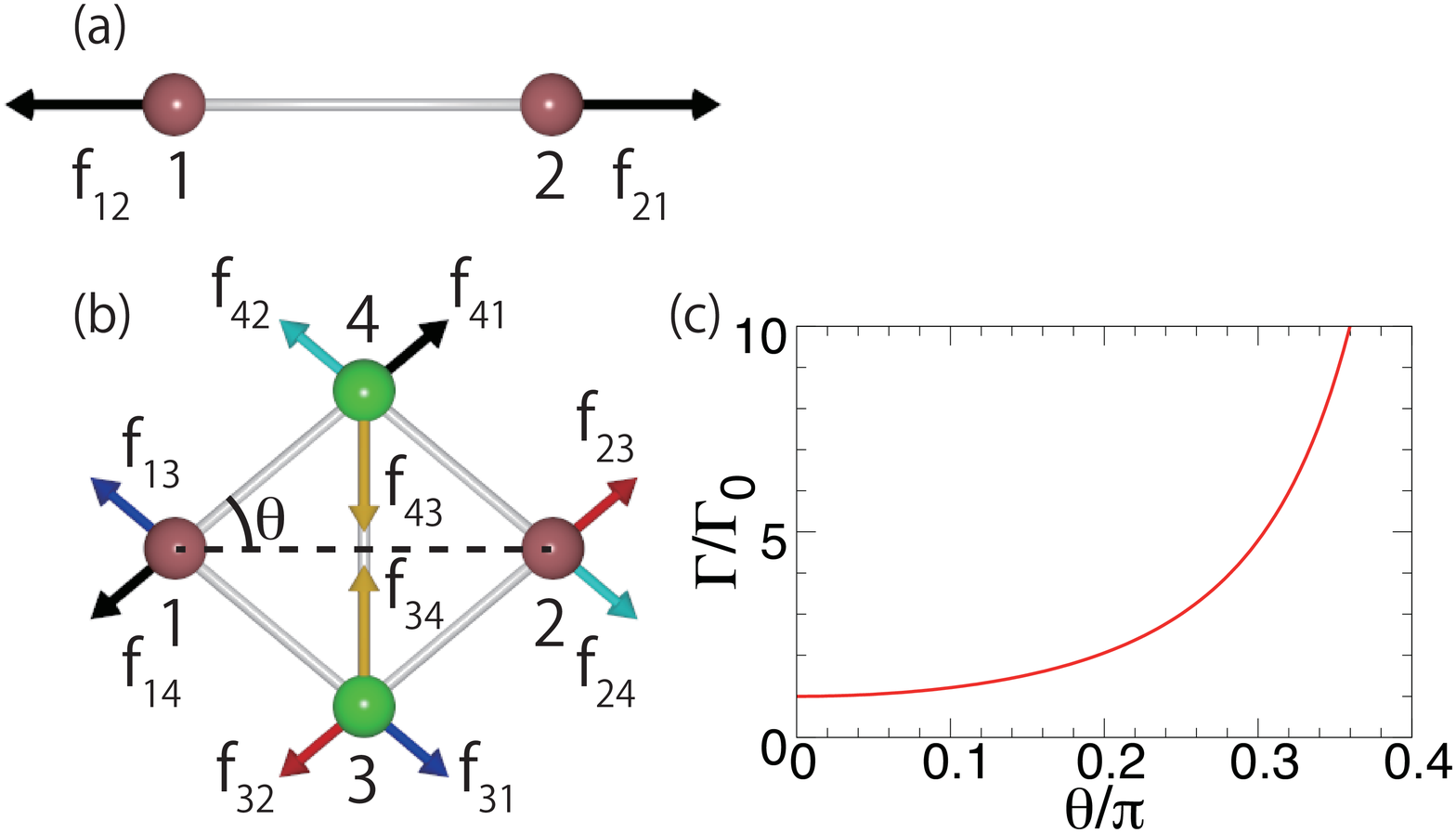}
  \caption{ 
    (Color online)
    Force decomposition for two-body forces.
    (a) Original forces. (b) Example of the hybrid decompositions.
    (c) Stress distribution magnitude $\Gamma$ of two-body forces as a function
    of $\theta$. It is normalized by the magnitude $\Gamma_0$ obtained by the IKN procedure.
  }
  \label{fig:pair_decomp}
\end{figure}

The hybrid decomposition can be applied to two-body forces
if two symmetric positions, ${\bf r}_{3}$ and ${\bf r}_{4}$, are employed 
 as shown in Fig.~\ref{fig:pair_decomp},
where $r_{14} = r_{13} = r_{24} = r_{23}$.
 Therefore, the IKN procedure is not a unique solution to obtain the stress tensor 
even for the two-body forces.
However,  the total length ($\ell_{\rm sum}=\sum_{i<j} r_{ij}$) and force norm sum ($f_{\rm sum}=\sum_{i<j} |f_{ij}|$) 
become greater than the IKN procedure.
Thus, the IKN procedure is the best decomposition method for two-body forces.

\subsection{Stress Distribution}\label{sec:stdis}

Although the force center can be set to an arbitrary position in HD,
positions that are excessively far away are not physically suitable.
Thus, we need a criterion to select the decomposition.
We consider the minimization of the stress distribution as a candidate criterion.
Hence, we define the stress-distribution magnitude $\Gamma$ as a summation over the cross norm of the stress,
\begin{equation}
\Gamma = \sum_{i<j} |f_{ij}|r_{ij},
\end{equation}
where the summation is taken over all pairs ($i,j \in {1, 2, 3, c}$ for the three-body forces).

For the two-body forces, 
the IKN procedure always gives the minimum value of $\Gamma$.
Therefore, $\Gamma$ can be employed as the criterion for the  two-body forces.
For the decomposition shown in Fig.~\ref{fig:pair_decomp}(b), 
the magnitude is given as $\Gamma = |f_{12}| r_{12} (1 + 2 \tan ^{2} \theta)$
and has the minimum at $\theta=0$ [see Fig.~\ref{fig:pair_decomp}(c)].

\begin{figure}
 \centering
 \includegraphics[width=60mm]{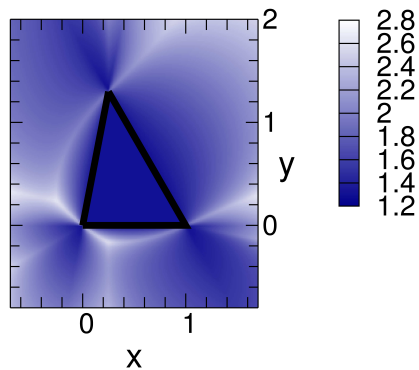}
 \caption{
(Color online)
Contour map of the stress distribution magnitude $\Gamma$ with respect to 
the force center position ${\bf r}_c(x,y)$ for the surface tension $k_{\rm {ar}}A_{123}$. 
The color bar shows the magnitude of $\Gamma/k_{\rm {ar}}$.}
 \label{fig:Gar}
\end{figure}

 In the following, we consider the minimization problem of $\Gamma$ for three-body forces.
For CFD and FCD, $\Gamma_{\rm {CFD}} = |f_{12}|r_{12}+|f_{13}|r_{13}+|f_{23}|r_{23}$
and $\Gamma_{\rm {FCD}} = |f_{1c}|r_{1c}+|f_{2c}|r_{2c}+|f_{3c}|r_{3c}$,
respectively.
Interestingly, when the original force center exists in the interior region of the triangle 
$\triangle 123$,
these two magnitudes take the same value: $\Gamma_{\rm {CFD}} =\Gamma_{\rm {FCD}}$.
The force norm sum $f_{\rm sum}$ of CFD is less than that of FCD,
while the total length $\ell_{\rm sum}$ of CFD is greater.
For HD with ${\bf r}_{c}$ lying in the interior region of $\triangle 123$,
\begin{eqnarray}
\Gamma_{\rm {HD}} &=& \Gamma_{\rm {HD}}^{\rm {FC}} +\Gamma_{\rm {HD}}^{\rm {CF}} \\ \nonumber
&=& (|f_{12}^{\rm {FC}}|+|f_{12}^{\rm {CF}}|)r_{12}+(|f_{13}^{\rm {FC}}|+|f_{13}^{\rm {CF}}|)r_{13} \\ && +(|f_{23}^{\rm {FC}}|+|f_{23}^{\rm {CF}}|)r_{23}.
\end{eqnarray}
When the CFD and FCD components in each force pair have the same sign,
{\it i.e.}, when $f_{12}^{\rm {CF}}f_{12}^{\rm {FC}}>0$, 
$|f_{12}^{\rm {all}}|= |f_{12}^{\rm {CF}}|+|f_{12}^{\rm {FC}}|$  so that  $\Gamma_{\rm {HD}} =\Gamma_{\rm {CFD}}$.
When $f_{12}^{\rm {CF}}f_{12}^{\rm {FC}}<0$,
$|f_{12}^{\rm {all}}|< |f_{12}^{\rm {CF}}|+|f_{12}^{\rm {FC}}|$ so that $\Gamma_{\rm {HD}} >\Gamma_{\rm {CFD}}$.
For the hybrid force center inside the triangle $\triangle 123$,
the decomposition with the same sign for each force pair can be chosen.
Thus, when ${\bf r}_c$ exists inside or on the edge of $\triangle 123$, 
$\Gamma$ takes the minimum value $\Gamma_{\rm {CFD}} =\Gamma_{\rm {FCD}}$.
Figure~\ref{fig:Gar} shows the minimum value of $\Gamma$ for each force center position ${\bf r}_c$ for an area potential.
Here, HD into three FCD force pairs and two CFD force pairs is used
($f_{12}^{\rm {CF}}=0$, $f_{13}^{\rm {CF}}=0$, or $f_{23}^{\rm {CF}}=0$), 
since infinitely small values can be taken for all FCD pairs if all six force pairs are allowed.
For ${\bf r}_{c}$ lying in the interior region of $\triangle 123$,
 $\Gamma$ is constant, while $\Gamma$ is greater for  ${\bf r}_{c}$ lying in the exterior region.
Therefore, the $\Gamma$ minimization implies the restriction on the decomposition choices to the interior region
but does not give a unique combination.

\begin{figure}
 \centering
 \includegraphics[width=80mm]{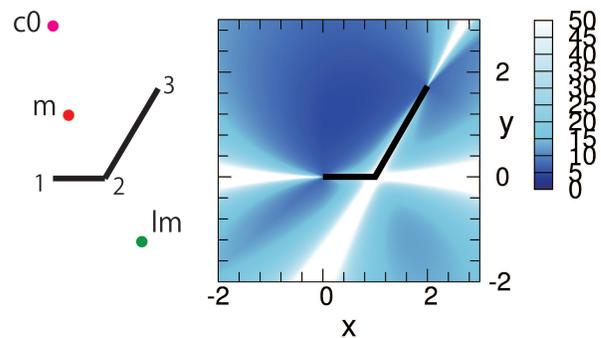}
 \caption{
(Color online)
Contour map of the stress distribution magnitude $\Gamma$ with respect to 
the force center position ${\bf r}_c(x,y)$ for the bending potential in Eq.~(\ref{eq:potbend1}). 
The color bar shows $\Gamma/k_{\rm b}$.
The positions of the original force center ${\bf r}_{c0}$ and 
the force centers ${\bf r}_{\rm m}$ and ${\bf r}_{\rm lm}$ of the global and local minima 
of $\Gamma$ are shown in the left panel.}
 \label{fig:Gben}
\end{figure}

When the original force center ${\bf r}_{c0}$ exists outside the triangle,
 $\Gamma$ typically has the lowest value at a single position of ${\bf r}_c$.
Figure~\ref{fig:Gben} shows  a typical example of $\Gamma$ in the HD of a bending potential on $\theta_{123}$ with $f_{13}^{\rm CF}=0$.
The deepest minimum of $\Gamma$ appears between ${\bf r}_{c0}$ and  the triangle $\triangle 123$ 
and local minima appear in the other exterior regions.
We consider the case where ${\bf r}_{c0}$ lies in the region B, as shown in Fig.~\ref{fig:q}.
We define the position ${\bf r}_{\rm m}$, which is geometrically determined:
\begin{eqnarray} \label{eq:rm}
{\bf r}_{\rm m} &=& {\bf r}_{\rm 2} + \sqrt{r_{12}r_{23}}\hat{{\bf r}}_{\rm {bv}}, \\
\hat{{\bf r}}_{\rm {bv}} &=& \frac{\hat{\bf r}_{12} + \hat{\bf r}_{32}}{|\hat{\bf r}_{12} + \hat{\bf r}_{32}|}, \nonumber
\end{eqnarray}
where $\hat{{\bf r}}_{\rm {bv}}$ is a unit vector bisecting the angle $\theta_{123}$.
The triangles $\triangle 12{\rm m}$ and $\triangle {\rm m}23$ are similar.
When ${\bf r}_{\rm m}$ is in the interior or on the edges of the triangle $\triangle 13c0$ formed by ${\bf r}_1$, ${\bf r}_3$, and ${\bf r}_{c0}$,
$\Gamma$ has the global minimum at ${\bf r}_{c}={\bf r}_{\rm {m}}$, 
where HD is taken for five force pairs, $f_{12}^{\rm CF}$, $f_{23}^{\rm CF}$, $f_{1m}$, $f_{2m}$, and $f_{3m}$.
This minimum appears not only for the bending forces but also for the other three-body forces with ${\bf r}_{c0}$ lying in the region B.
The local minimum in the region E with $f_{13}^{\rm {CF}}=0$ appears at ${\bf r}_{\rm lm}= {\bf r}_{\rm 2} - \sqrt{r_{12}r_{23}}\hat{{\bf r}}_{\rm {bv}}$.
The derivations of these global and local minima are described in Appendix~\ref{sec:Gmin}.
The stress cross norms are balanced at ${\bf r}_{\rm m}$: $|f_{1m}|r_{1{\rm m}}=|f_{2m}|r_{2{\rm m}}=|f_{3m}|r_{3{\rm m}}$. 
For the bending forces,
the condition for ${\bf r}_{\rm {m}}$ lying in $\triangle 13c0$
is $r_{12}/r_{23} \ge \cos^2(\theta_{123}/2)$ and $r_{23}/r_{12} \ge \cos^2(\theta_{123}/2)$.
This condition is satisfied in typical simulation conditions including our present simulation.
It is violated only when $r_{12}/r_{23}$ significantly deviates from unity and $\theta_{123}$ is small.

For general three-body forces,
 ${\bf r}_{\rm {m}}$ can be outside $\triangle 13c0$.
In this case, we do not have an analytical solution for the $\Gamma$ minimum,
but it can be calculated numerically.
In the next section, we investigate how the stress profile of a bilayer membrane depends on the decomposition.

\begin{table}
 \begin{center}
  \begin{tabular}{c|r|r|r} \hline
    & W\ \  & H\ \ & T\ \ \\ \hline
  W & 25 & 25 &\ \ 200 \\
  H & 25 & 25 &\ \ 200 \\
  T &\ \ 200&\ \ 200 & 25 \\ \hline
  \end{tabular}
  \caption{Repulsive interaction parameters $a_{ij}$ with unit $k_{\mathrm B}T$. \label{tab:intra}}
 \end{center}
\end{table}

\section{Bilayer membrane}\label{sec:mem}

We simulate a tensionless bilayer membrane with various
decompositions of bending forces using coarse-grained and atomistic lipid models.
In Sec.~\ref{sec:cg_model}, the stress profile and Gaussian curvature modulus
are discussed using  the dissipative particle dynamics (DPD)
method~\cite{groot1997dissipative,venturoli2006mesoscopic,hoogerbrugge1992simulating,espanol1995statistical}.
DPD is one of the widely used coarse-grained lipid models.
In Sec.~\ref{sec:atom_model}, the stress profile of
 an atomistic MD of DOPC (1,2-Dioleoyl-sn-glycero-3-phosphocholine) using CHARMM36 force field~\cite{klauda2012improving,klauda2010update} is discussed.

We refer to HD with the global and local minima of $\Gamma$ in the regions B and E as HD(GM) and HD(LM), respectively.
In HD, we examine only the case $f_{13}^{\rm CF}=0$, since HD(GM) and HD(LM) are obtained in this condition.

\subsection{Coarse-grained model}\label{sec:cg_model}

\subsubsection{Model description}\label{sec:model}
An amphiphilic molecule is represented by a linear chain of four particles: one hydrophilic (H) and three hydrophobic (T) DPD particles.
Neighboring DPD particles are connected via the harmonic bond 
potential,
$U_{\mathrm{bond}}(r_{ij}) = (k_{\mathrm {s}}/2)(1 - r_{ij}/\ell_0)^{2}$,
with $k_{s} = 150k_{\mathrm B}T$, where $k_{\mathrm B}T$ is the thermal energy.
One of the simplest bending potentials is employed at the second and third particles of the amphiphile:
\begin{equation} \label{eq:potbend1} 
U_{\mathrm{bend1}}(\theta_{ijk})= k_{\rm b}(1 - \cos \theta_{ijk} ),
\end{equation}
with $k_{\rm b} = 30 k_{\mathrm B}T$.
A dihedral potential is not considered.
Water is represented by DPD particles labeled W.
All particle pairs interact through a soft repulsive potential:
$U_{\mathrm{rep}}(r_{ij}) = (a_{ij}/2) (1 - r_{ij}/r_{\rm {cut}})^{2}$,
which vanishes beyond the cutoff at $r_{ij} = r_{\rm {cut}}$.
We set $r_{\rm {cut}}=2\ell_0$ in this study.
The repulsive interaction parameters, $a_{ij}$, are listed in Table~\ref{tab:intra}. 

The amphiphilic molecules form a bilayer membrane with the bending rigidity $\kappa/k_{\mathrm B}T = 18.3 \pm 0.2$, 
which is a typical value for a bilayer membrane at room temperature~\cite{lipowsky1995structure}.
The details of the simulation method are described in Appendix~\ref{sec:simcg}.

%% Lateral pressure profile for different force decomposition.
\subsubsection{Lateral pressure profile}\label{sec:press}

\begin{figure}
 \centering
 \includegraphics[width=80mm]{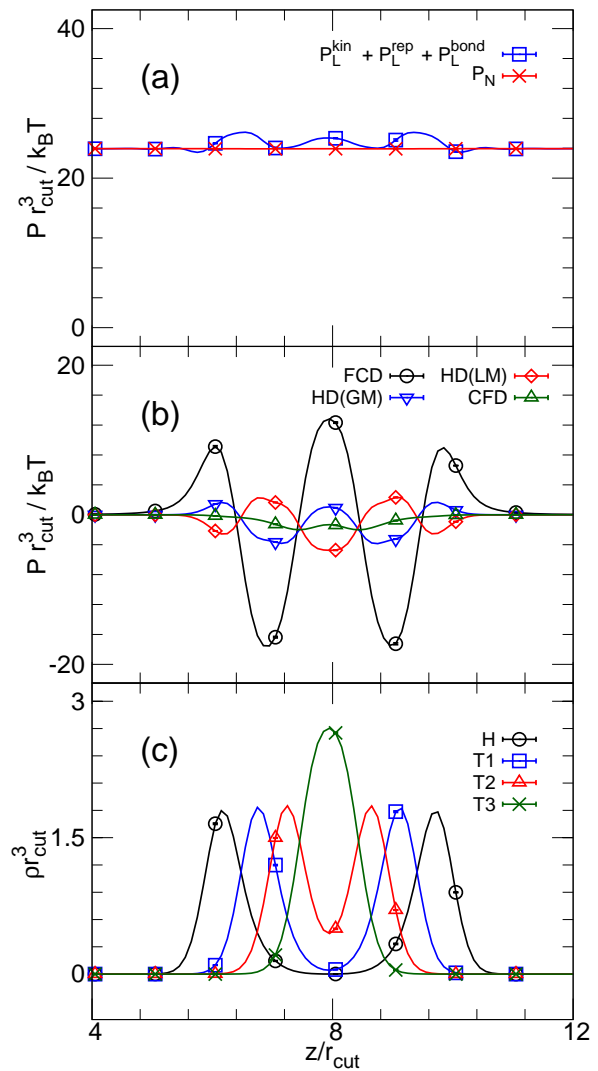}
 \caption{
(Color online)
Pressure and density profiles along bilayer normal ($z$) axis 
obtained by the DPD simulation.
(a)
Total normal pressure profile  $P_{\rm N}(z)$  and
partial lateral pressure profile $P_{\rm L}(z)$ given by the sum of three contributions of the kinetic, repulsive, and bond potential components.
(b) 
Lateral pressure profile $P_{\rm L}^{\rm bend1}(z)$ given by bending potential stress with four decomposition methods.
(c) Number density profile of four particles in the amphiphilic molecules.
H represents the first (hydrophilic head) particle.
T1, T2, and T3 represent three hydrophobic particles.
The symbols and error bars are shown at several data points. 
}
 \label{fig:loc_stress_repbond}
\end{figure}

%%% Local stress strongly depends on force decomposition methods.
The lateral and normal pressure profiles along the normal ($z$) direction of the bilayer membrane
for different force decomposition methods are shown in Fig.~\ref{fig:loc_stress_repbond}.
The pressure profiles are calculated from the average stress for small slices along the $xy$ plane with a width of $\Delta z = 0.2\ell_0$:
 $P_{\rm L}(z) = -(\sigma_{xx}(z)+\sigma_{yy}(z))/2$ and $P_{\rm N}(z) = -\sigma_{zz}(z)$.
The lateral profile $P_{\rm L}(z)$ strongly depends on the force decomposition methods,
while the normal profile $P_{\rm N}(z)$ is independent of the decompositions and takes a constant value.
The contribution of two-body forces to $P_{\rm L}(z)$ is only slightly dependent on $z$ [see Fig.~\ref{fig:loc_stress_repbond}(a)].

The contribution $P^{\rm{bend1}}_{\rm L}$ of the bending forces to the
lateral profile is significantly different for different decomposition methods.
The amplitude of $P^{\rm{bend1}}_{\rm L}$ of FCD is much larger than those of CFD, HD(GM), and HD(LM),
as shown in Fig.~\ref{fig:loc_stress_repbond}(b).
Surprisingly, the function shape of $P_{\rm L}$ calculated by HD(LM) has the opposite sign
those calculated by the other force decomposition methods.
In addition, the pressure peaks of FCD slightly shift to the outside of the position of the head particles of the bilayer
[compare Figs.~\ref{fig:loc_stress_repbond}(b) and (c)].
As mentioned in the previous section, 
for all force decompositions shown in Fig.~\ref{fig:loc_stress_repbond}, linear- and angular-momentum conservation are satisfied.

\begin{figure}
 \centering
 \includegraphics[width=80mm]{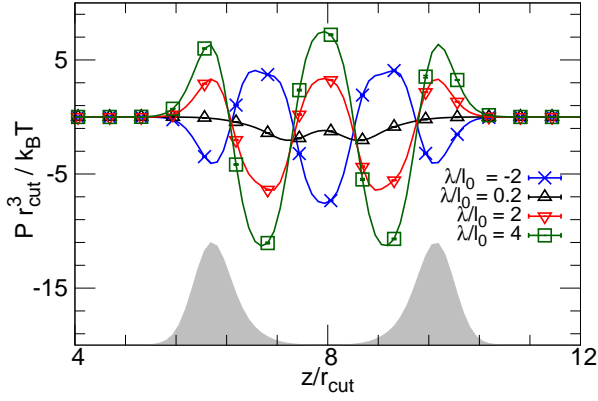}
 \caption{
(Color online)
Length $\lambda$ dependence of the bending potential contribution $P^{\rm{bend1}}_{\rm L}$ to the lateral pressure profile.
The symbols and error bars are shown at several data points. 
The density profile of hydrophilic heads [the same data in Fig.~\ref{fig:loc_stress_repbond}(c)]  is shown as the gray-filled curve in arbitrary units.
}
 \label{fig:loc_stress_lambda}
\end{figure}

\begin{figure}
 \centering
 \includegraphics[width=80mm]{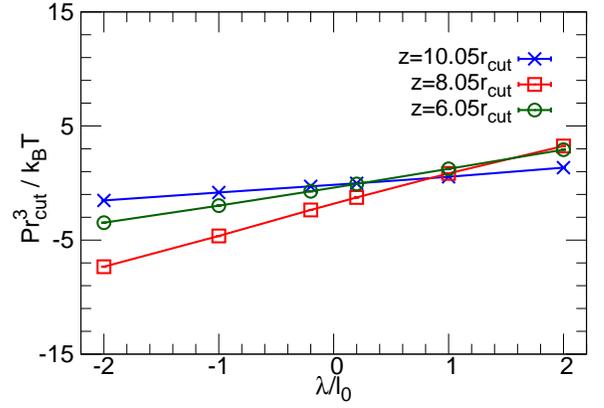}
 \caption{
(Color online)
Contribution of bending potential $P^{\rm{bend1}}_{\rm L}$ to the lateral pressure
 at three $xy$ planes with $z/r_{\rm {cut}}=6.05$, $8.05$, and $10.05$ as a function of $\lambda$.}
 \label{fig:press_vs_lambda}
\end{figure}

%%% Lateral pressure profile dependency of lambda.
To further examine the dependence of lateral pressure on the force decomposition,
we systematically change the force center ${\bf r}_c$:
\begin{align}
 {\bf r}_c = {\bf r}_{2} + \lambda \hat{{\bf r}}_{\mathrm{bv}},
\end{align}
where $\lambda$ is the distance between ${\bf r}_c$ and ${\bf r}_{2}$.
For HD(GM) and HD(LM), $\lambda=\sqrt{r_{12}r_{23}}$ and  $\lambda=-\sqrt{r_{12}r_{23}}$, respectively.
At  $\lambda=2r_{12}r_{23}\cos (\theta_{123}/2)/(r_{12}+r_{23})$,
the decomposition corresponds to CFD, since the force center is on the line segment between ${\bf r}_{1}$ and  ${\bf r}_{3}$.
Figures~\ref{fig:loc_stress_lambda} and \ref{fig:press_vs_lambda} show
the dependence of $P^{\rm{bend1}}_{\rm L}$ on $\lambda$.
As $\lambda$ increases, the lateral pressure increases.
A linear relation between $\lambda$ and $P^{\rm{bend1}}_{\rm L}$ (also $\lambda$ and $P_{\rm L}$) 
is found even for negative values of $\lambda$.

%%% show there exists the linear relationship between maximum pressure and lambda by exemplifying simple molecular model.
This linear dependence on $\lambda$ is analytically derived 
when $\hat{{\bf r}}_{\mathrm{bv}}$ is along the $x$ axis and $\hat{\bf r}_{12}-\hat{\bf r}_{32}$ is along the $z$ axis.
Since the force pair ${\bf f}_{2c}$ contributes to the stress $\sigma_{xx}(z)$ as $f_{2c}\lambda\delta(z-z_2)/A_{xy}$,
the lateral stress produced by the bending potential on $\theta_{123}$ 
is given by
\begin{align}
\sigma_{xx}^{\rm {bend}}(z) =& \frac{f_{1}r_{12}}{A_{xy} \sin \varphi}\Big[
 s_{\rm {b1}}B( z_1, z_2, z) \nonumber \\
&+  s_{\rm{b2}}\delta(z-z_2) 
+  s_{\rm {b3}}B( z_3, z_2, z) \Big] ,  \label{eq:delta_pz} \\
 s_{\rm {b1}} =& -\frac{\lambda}{r_{12}} + 2\cos \varphi - \cos^{3} \varphi , \nonumber \\
 s_{\rm{b2}}  =&  \frac{r_{12}+r_{23}}{r_{12}r_{23}}\lambda - 2\cos \varphi , \nonumber \\
 s_{\rm {b3}} =& - \frac{\lambda}{r_{23}} + 2\cos \varphi - \cos^{3} \varphi , \nonumber 
\end{align}  \noindent
where $\varphi=\theta_{123}/2$ and $A_{xy}$ is the area of the $xy$ plane.
Equation~(\ref{eq:delta_pz}) clearly shows that $\sigma_{xx}^{\rm {bend}}(z)$ is a linear function of $\lambda$ for $z_{1} < z < z_{3}$.
Our simulation results indicate that this linear relation is approximately satisfied even when averaging
the conformations in which $\hat{{\bf r}}_{\mathrm{bv}}$ are fluctuated around the $xy$ plane.

\subsubsection{Gaussian curvature modulus}\label{sec:gauss}
%% Verification of lateral stress profile.
%%% Calculation of Gaussian curvature modulus.
The Gaussian curvature modulus $\bar{\kappa}$ can be calculated~\cite{safran2003statistical,helfrich1994lyotropic,hu2013gaussian} as 
\begin{align}
 \bar{\kappa} = \int \{P_{\rm N}(z)- P_{\rm L}(z)\} z^{2} dz.
 \label{eq:moment}
\end{align}
From elastic theory, $\bar{\kappa}$ is related with $\kappa$ via~\cite{landau1986theory}
\begin{align}
 \bar{\kappa} = (\nu - 1)\kappa,
 \label{eq:bar_kappa_kappa}
\end{align}
where $\nu$ is the Poisson's ratio of the bilayer membrane.
Though the Poisson's ratio is generally varied in the range of $-1\le \nu \le 1/2$,
$\bar{\kappa}/\kappa \simeq -1$ was reported in the simulations by Hu {\it et al}.~\cite{hu2013gaussian,hu2012determining} and experiments~\cite{baumgart2005membrane,semrau2008accurate}. 
Hu {\it et al}. calculated $\bar{\kappa}$ from the shape transition between a disk-shaped bilayer patch and vesicle.
They also calculated  $\bar{\kappa}$ using the pressure profile with
Eq.~(\ref{eq:moment}) but concluded that the pressure profile yields
unphysical results since the resultant $\bar{\kappa}$ is positive or
has a small amplitude compared to $\kappa$.
However, their pressure-profile calculation was performed using GLD; hence, the pressure tensor does not satisfy angular-momentum conservation.
Recently, Torres-S{\'a}nchez {\it et al}. calculated $\bar{\kappa}$ using CFD~\cite{torres2015examining}.
They reported that the calculated $\bar{\kappa}$ agrees well with experimental values.

\begin{figure}
 \centering
 \includegraphics[width=80mm]{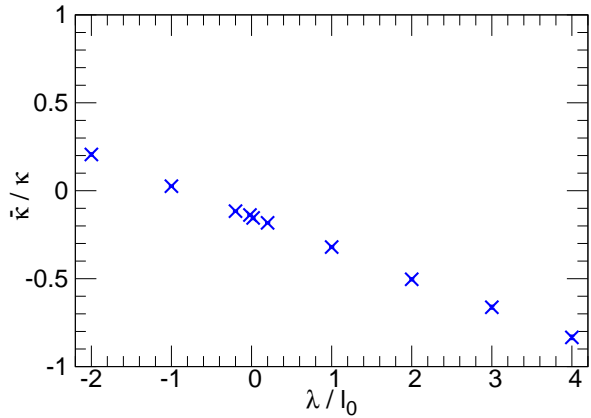}
 \caption{
(Color online)
Ratio of Gaussian curvature modulus $\bar{\kappa}$ to bending rigidity $\kappa$ as a function of $\lambda$.}
 \label{fig:kappa_vs_lambda}
\end{figure}

As described in Sec.~\ref{sec:press}, 
the lateral pressure profile is strongly dependent on the force decomposition method.
Thus, $\bar{\kappa}$ estimated with Eq.~(\ref{eq:moment})
also varies significantly on changing the force center in HD.
Table~\ref{tab:kappa_bars} lists $\bar{\kappa}$ and $\bar{\kappa}/\kappa$ for four different decomposition methods.
CFD, HD(GM), and HD(LM) give  $-\bar{\kappa}/\kappa \ll 1$,
and FCD gives $-\bar{\kappa}/\kappa > 1$.
None of them satisfy $\bar{\kappa}/\kappa \simeq -1$.
To further clarify the dependence of $\bar{\kappa}$ on ${\bf r}_{c}$,
we calculated $\bar{\kappa}/\kappa$ as a function of $\lambda$. 
Figure~\ref{fig:kappa_vs_lambda} shows the linear dependence of $\bar{\kappa}/\kappa$ on $\lambda$. 
This linearity is the consequence of the linearity of the pressure profile on $\lambda$.
When $\lambda \simeq 4\ell_0$,
$\bar{\kappa}/\kappa \simeq -1$ is obtained.
However, this position is too far from the positions of the interacting particles.
Thus, it does not seem to be physically plausible.
Our results support Hu's conclusion that  Eq.~(\ref{eq:moment}) gives
an unphysical value of $\bar{\kappa}$ in bilayer membranes.

\begin{table}
 \begin{center}
  \begin{tabular}{c|c|c} \hline
    & $\bar{\kappa}/k_{\mathrm B}T $ & $\bar{\kappa} / \kappa$ \\ \hline
   CFD  & -3.1 $\pm$ 0.2 & -0.17 $\pm$ 0.01 \\
   HD(GM)  & -6.15 $\pm$ 0.09 & -0.335 $\pm$ 0.006 \\
   HD(LM) & 0.72 $\pm$ 0.07 & 0.039 $\pm$ 0.004 \\
   FCD & -32.8 $\pm$ 0.1 & -1.79 $\pm$ 0.02 \\ \hline
   \end{tabular}
  \caption{\label{tab:kappa_bars}
Gaussian curvature modulus $\bar{\kappa}$ and its ratio to bending rigidity $\bar{\kappa}/\kappa$ for different force decomposition methods.}
 \end{center}
\end{table}

\subsection{Atomistic model}\label{sec:atom_model}
\subsubsection{Model description}
The DOPC molecules are modeled by the recent version of CHARMM all-atom force field
(CHARMM36)~\cite{klauda2012improving,klauda2010update},
and water molecules are modeled by rigid TIP3P. We apply CFD,
FCD, HD(LM), and HD(GM) to the bending potential.
The four-body potential
contribution to local stress field is calculated using CFD.
The details of the simulation method are described in Appendix~\ref{sec:simatom}.

\subsubsection{Lateral pressure profile}
\begin{figure}
 \centering
 \includegraphics[width=80mm]{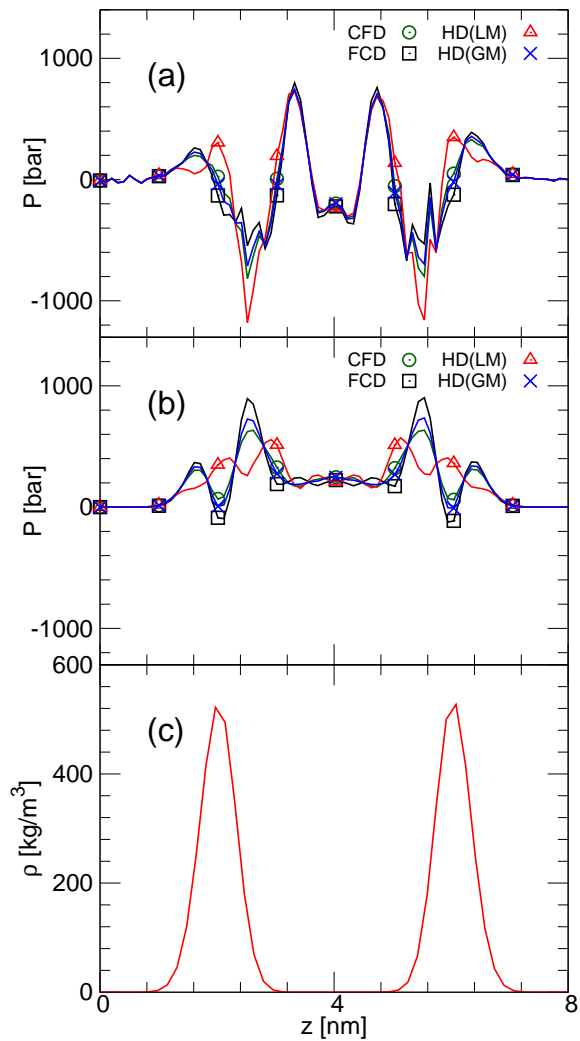}
 \caption{
   (Color online) Pressure and density profiles of DOPC membrane along the bilayer normal ($z$) axis.
(a) Total lateral pressure profile for
   four force decomposition methods. (b) Partial lateral pressure profile
   given by the bending potential. (c) Mass
   density profile of hydrophilic head groups (phosphoric acid and choline groups).
 }
 \label{fig:stress_atomis}
\end{figure}

The lateral pressure profiles along the bilayer normal direction are shown in
Fig.~\ref{fig:stress_atomis} for four different force decomposition methods.
The pressure profiles are calculated for small slices with slice width $\Delta z=0.1$nm in
the same manner as in Sec.~\ref{sec:cg_model}.
The dependence of lateral pressure profile on the force
decompositions is qualitatively similar to that of the DPD model
but its amplitude becomes much smaller [see Fig.~\ref{fig:stress_atomis}(b)].  
The differences of force
decompositions affect the local pressure at the surface between water and
amphiphilic molecules. 
In the hydrophobic region, there are no significant
differences of stress profiles for different force decompositions.
Thus, in the higher-resolution model, the decomposition methods of the bending forces
modify the pressure profile less than the lower-resolution (coarse-grained) model.

\section{Discussion}\label{sec:dis}

The total stress of each three-body potential is independent of the decomposition method.
However, the distribution of this stress in 2D space significantly varies even under the strong law of action and reaction.
We introduced the stress distribution magnitude $\Gamma$ to evaluate the decomposition method.
For the area potentials, $\Gamma$ has the minimum value in the entire
triangular region formed by the three particles,
whereas $\Gamma$ has the minimum value at a single position for the bending potentials.
Hence, $\Gamma$ can be used to reduce the candidates for suitable decompositions but
the best (unique) decomposition is not determined by $\Gamma$, at least for the area potential.

The discrete stress of a molecular simulation can be mapped into the stress field in the continuum space.
If the corresponding stress field in the continuum space is known,
one can state that the decomposition producing the closest stress is the best choice.
In typical simulation conditions, the resultant stress cannot be obtained a priori.
However, if a particle potential is constructed as a discretized version of the potential in the continuum space,
the corresponding stress field  in the continuum space is obtained from the original continuum potential.
The surface tension $k_{\rm ar}A_{123}$ is one of the  discretized potentials.
When a continuum surface with area $A$ is discretized to acute triangles,
the surface tension of $k_{\rm ar}A$ 
is discretized to  $k_{\rm ar}\sum_k A_{k}$, where $A_k$ is the area of the $k$-th triangle.
When the triangle is on the $xy$ plane,
$\sigma_{xx}({\bf x})=\sigma_{yy}({\bf x})=k_{\rm ar}/L_z$ and $\sigma_{xy}({\bf x})=0$ are given in the continuum description,
where $L_z$ is the side length of the simulation box in the $z$ direction.
Both CFD and FCD distribute the stress into line segments so that they deviate from the constant stress field.
If HD with multiple force centers distributed on the triangle is employed,
a nearly constant stress field can be constructed.
Alternatively, Hardy's spatial average with a weighting function~\cite{hardy1982formulas} also helps CFD and FCD  to approach the constant field.

For surface tension or other discretized potentials,
the resultant stress field becomes closer to the original continuum field  as the surface is discretized into smaller triangles.
Thus, it is related with the resolution of the simulation.
For classical molecular simulations, local interactions in a length scale smaller than the diameter of atoms or particles
are not typically taken into account for coarse-graining.
For  all-atom simulations, the force fields between atoms are
constructed from ab initio quantum mechanical
calculations~\cite{mackerell2004empirical, cornell1995second, mackerell1998all}.
Even from the viewpoint of classical mechanics, each particle has a finite size.
For a pairwise interaction such as chemical bonds, the stress is distributed not only in the line segment between two particle centers
but also in a cylindrical region with the diameter equal to the particle size.
Thus, one may have to determine the decomposition method for multibody forces through comparison with the underlying high-resolution potential interactions.
For lipid membranes,
the pressure profile of the higher-resolution atomistic model 
has much smaller dependence on the decomposition than that of the lower-resolution (four-particle) DPD model.
This also supports our hypothesis on the resolution.

Let us go back to the discussion on the stress field of the bending forces on $\theta_{123}$.
CFD and HD(GM) of the bending forces give the stress distribution
on the edge of ${\bf r}_{13}$ or  close to the edge, respectively.
If these positions are within the interaction radius of the atom (or particle) at ${\bf r}_{1}$ or ${\bf r}_{3}$,
they can be employed as a force-acting point.
However, 
FCD and HD(LM) are unphysical since their force centers 
are far from the triangle $\triangle 123$ in most of the case. 
For real bending potentials, the stress distribution may strongly depend
on the molecules,
but it is likely approximated to the interaction between two chemical bonds (the middle points of ${\bf r}_{12}$ and ${\bf r}_{23}$).
Thus, HD with the force center ${\bf r}_{c}$ lying in the middle of $\triangle 123$
may be a physically reasonable decomposition, 
where $\Gamma$ is greater than those of HD(GM) and CFD but the stress profile of the bilayer membranes is flatter.

A coarse-grained model often does not have a specific underlying higher-resolution model.
In such a case, one may have to calculate the stress field without the higher-resolution information.
We describe our speculative consideration on the choice of the decomposition when
the force center ${\bf r}_{c0}$ 
lies in the interior region of the triangle of three interacting particles like in the area potential.
In this case, FCD, CFD, and HD which force center lying in the interior of the triangle
has the minimum value of $\Gamma$.
Among of them,
FCD gives the minimum of the total length $\ell_{\rm sum}$, {\it i.e.},
the minimum propagation path of the stress.
Therefore, the minimum total length may be employed as an additional criterion
so that FCD can be chosen.

For $n$-body forces with $n\ge 4$, all of the extrapolations of the
force vectors ${\bf f}_i$ from ${\bf r}_i$ do not typically meet at a single position.
Thus, FCD is not generally  available for $n\ge 4$.
However, FCD can be performed for specific potentials for which all ${\bf f}_i$ meet a single position.
Let us consider potentials $U_{\rm {rg}}(r_{\rm gw}^2)$ on a weighted radius of gyration 
$r_{\rm gw}^2=\sum_i^n w_i({\bf r}_i-{\bf r}_{\rm Gw})^2$ for a center position  
${\bf r}_{\rm Gw}=\sum_i^n w_i{\bf r}_i$, where the weight $w_i$ is normalized as $\sum_i^n w_i=1$.
Since all of ${\bf f}_i$ meet at ${\bf r}_{\rm Gw}$, these forces are decomposed by FCD with the force center ${\bf r}_{c}={\bf r}_{\rm Gw}$.
If the force center ${\bf r}_{c}={\bf r}_{\rm Gw}$ or  a force center for three of the forces is used,
HD is applicable for any $n$-body force.
The center position ${\bf r}_{c}$ can be arbitrarily set by adjusting $w_i$ in ${\bf r}_{\rm Gw}$.
Multiple center positions may be useful.
However, it has many choices of the force decomposition for $n\ge 4$,
and it is currently unclear how the force decomposition can be tuned.

\section{summary}\label{sec:sum}

We have proposed a decomposition method (FCD) of three-body forces using the position, where three force extrapolations from the particle positions meet,
and combined it with CFD, which decomposes the forces into force pairs between interacting particles.
Our study has revealed that the local stress field of three-body forces
is strongly dependent on these decomposition methods.
We have discussed the choice of the decomposition using the stress distribution magnitude $\Gamma$ and comparison with
the stress fields in continuum fields and in higher resolutions of discretization.
We have not reached a concrete conclusion for the best decomposition 
but rather considered that it depends on the underlying higher-resolution potential.

\begin{acknowledgments}
This work was supported by JSPS KAKENHI Grant Number JP25103010 and JP16J01728.
\end{acknowledgments}

\begin{appendix}

\section{Area Potential}\label{sec:area}

Here, we describe the force decomposition of the general form of area potentials, $U_{\rm {ar}}({A_{123}})$, 
for the triangle $\triangle 123$.
The area  is given by $A_{123}=|{\bf r}_{12}\times {\bf r}_{13}|/2$.
The force ${\bf f}_1$ is given as 
\begin{equation}
{\bf f}_{1} =  -\frac{\partial U_{\rm {ar}}}{\partial {\bf r}_{1}} 
=  - \frac{U'_{\rm {ar}}}{4A_{123}}[r_{23}^2{\bf r}_{13} - ({\bf r}_{13}\cdot{\bf r}_{23}){\bf r}_{23}],
\end{equation}
where $U'_{\rm {ar}}=\partial U_{\rm {ar}}/\partial {A_{123}}$.
This force ${\bf f}_{1}$ is perpendicular to ${\bf r}_{23}$,
since the area does not change if  ${\bf r}_1$  moves parallel to ${\bf r}_{23}$.
For the potential of the surface tension $U_{\rm {ar}}({A_{123}})=k_{\rm {ar}} A_{123}$,
$U'_{\rm {ar}}=k_{\rm {ar}}$.

The force $f_{12}$ in CFD is obtained  by the decomposition of ${\bf f}_1$ into components along $\hat{\bf r}_{12}$ and  $\hat{\bf r}_{13}$
or directly by using $f_{12}=-\partial U_{\rm {ar}}/\partial r_{12}|r_{13},r_{23}$ 
with Heron's formula $A_{123}=\sqrt{b(b-r_{12})(b-r_{13})(b-r_{23})}$,
where $b=(r_{12}+r_{13}+r_{23})/2$:
\begin{equation}
f_{12} = - \frac{U'_{\rm {ar}}}{4A_{123}}({\bf r}_{13}\cdot{\bf r}_{23})r_{12}.
\end{equation}
The other forces ${\bf f}_2$, ${\bf f}_3$, $f_{13}$, and $f_{23}$ are similarly obtained.
The original force center ${\bf r}_{c0}$ is the orthocenter of $\triangle 123$. 
Since $q=1/4{U'_{\rm {ar}}}^2>0$,
 ${\bf r}_{c0}$ lies in the interior region 
or exterior region A, C, or E of $\triangle 123$ depicted in Fig.~\ref{fig:q}.
When $\triangle 123$ is an acute triangle, 
${\bf r}_{c0}$ lies in the interior region.
When the angle $\theta_{123}$ is obtuse (${\bf r}_{12}\cdot {\bf r}_{32}<0$),
${\bf r}_{c0}$ lies in the exterior region E.

\section{Bending Potential}\label{sec:bend}

Next, we describe the force decomposition of the general form of bending potentials, 
$U_{\rm {bend}}(\hat{\bf r}_{12}\cdot\hat{\bf r}_{32})$, 
for the angle $\theta_{123}=\cos^{-1}(\hat{\bf r}_{12}\cdot\hat{\bf r}_{32})$ of 
three particle positions ${\bf r}_1$, ${\bf r}_2$, and ${\bf r}_3$.
The forces on the three particles are given by
\begin{eqnarray} \label{eq:benf}
{\bf f}_{1} &=&  - \frac{U'_{\rm {bend}}}{r_{12}}[\hat{\bf r}_{32} - (\hat{\bf r}_{12}\cdot\hat{\bf r}_{32})\hat{\bf r}_{12}], \\ 
{\bf f}_{2} &=&  - \frac{U'_{\rm {bend}}}{r_{12}r_{32}}\Bigg[ \frac{({\bf r}_{12}\cdot{\bf r}_{32} - r_{12}^2)\hat{\bf r}_{12}}{r_{12}} 
+ \frac{({\bf r}_{12}\cdot{\bf r}_{32} - r_{32}^2)\hat{\bf r}_{32}}{r_{32}}\Bigg] , \nonumber \\ \nonumber
{\bf f}_{3} &=&  - \frac{U'_{\rm {bend}}}{r_{32}}[\hat{\bf r}_{12} - (\hat{\bf r}_{12}\cdot\hat{\bf r}_{32})\hat{\bf r}_{32}]. 
\end{eqnarray}
The forces ${\bf f}_{1}$ and ${\bf f}_{3}$ are perpendicular to ${\bf r}_{12}$ and ${\bf r}_{32}$, respectively,
since $\theta_{123}$ is independent of the lengths $r_{12}$ and $r_{32}$.
For the bending potential of Eq.~(\ref{eq:potbend1}),
$U'_{\rm {bend}}=-k_{\rm b}$.

In CFD, these forces are decomposed into the following force pairs:
\begin{eqnarray} \label{eq:bencfd}
f_{12} &=&  - U'_{\rm {bend}}\frac{ {\bf r}_{12}\cdot{\bf r}_{13}}{r_{12}^2r_{23}}  ,  \nonumber \\
f_{13} &=& \ \   U'_{\rm {bend}}\frac{r_{13}}{r_{12}r_{23}}  , \\  \nonumber
f_{23} &=&  - U'_{\rm {bend}}\frac{{\bf r}_{23}\cdot{\bf r}_{13}}{r_{12}r_{23}^2} .
\end{eqnarray}
These force pairs can be obtained from Eqs.~(\ref{eq:benf}) and (\ref{eq:f12a}) or 
directly from $f_{12}=-\partial U_{\rm {bend}}/\partial r_{12}|r_{13},r_{23}$
 with $\hat{\bf r}_{12}\cdot\hat{\bf r}_{32}= (r_{12}^2+r_{23}^2-r_{13}^2)/2r_{12}r_{23}$.
The original force center ${\bf r}_{c0}$ always lies in the exterior region  of $\triangle 123$,
since $q=-4A_{123}^2/r_{12}^4 r_{23}^4 {U'_{\rm {bend}}}^{2}<0$.
When the angles $\theta_{312}<\pi/2$ and $\theta_{231}<\pi/2$,
$f_{12}f_{23}>0$ and $f_{12}f_{13}<0$ so that
${\bf r}_{c0}$ lies in the exterior region B depicted in Fig.~\ref{fig:q}.
For $\theta_{312}>\pi/2$ or $\theta_{231}>\pi/2$,
${\bf r}_{c0}$ lies in the region D or F, respectively.
The stress distribution magnitudes $\Gamma$ for FCD and CFD
take the same value for the bending potentials: 
$\Gamma_{\rm {FCD}}=\Gamma_{\rm {CFD}}=2r_{13}^2|U'_{\rm {bend}}|/r_{12}r_{23}$ for $\theta_{312}<\pi/2$ and $\theta_{231}<\pi/2$,
and 
$\Gamma_{\rm {FCD}}=\Gamma_{\rm {CFD}}=2{\bf r}_{12}\cdot{\bf r}_{13}|U'_{\rm {bend}}|/r_{12}r_{23}$ for $\theta_{231}>\pi/2$.
For the typical simulation conditions including our present simulation,
$\theta_{312}$ and $\theta_{231}$ are small.
Thus, we consider only the case of ${\bf r}_{c0}$ lying in region B in this paper.

\section{Minimization of Stress Distribution Magnitude for Exterior Force Center}\label{sec:Gmin}

Here, we consider the force center ${\bf r}_{c}$  for the minimum of the stress distribution magnitude $\Gamma$, 
when ${\bf r}_{c0}$ lies in the exterior region B, where $f_{12}f_{13}<0$, $f_{12}f_{23}>0$, and $q<0$.
As mentioned in  Sec.~\ref{sec:fcd},
$\Gamma$ takes the lowest value at the position ${\bf r}_{\rm m}$ given in Eq.~(\ref{eq:rm})
for HD with $f_{13}^{\rm {CFD}}=0$,
if ${\bf r}_{\rm m}$ is in the interior region surrounded by three positions ${\bf r}_1$, ${\bf r}_3$, and ${\bf r}_{c0}$.
This position is derived as follows.
We consider the minimization of the difference $\Gamma_{\rm {dif}}=\Gamma - \Gamma_{\rm {CFD}}=\Gamma_{\rm {HD}}^{\rm {FC}}- (|f_{12}^{\rm {FC}}|r_{12}+|f_{13}^{\rm {FC}}|r_{13}+|f_{23}^{\rm {FC}}|r_{23})$,
since the contribution of the CFD force pairs does not explicitly appear in $\Gamma_{\rm {dif}}$.
\begin{eqnarray}
\Gamma_{\rm {dif}} &=& \frac{2\tilde{f}_{12}^{\rm {FC}}\tilde{f}_{23}^{\rm {FC}}}{|q|}(|\tilde{f}_{12}^{\rm {FC}}|r_{12}^2+|\tilde{f}_{23}^{\rm {FC}}|r_{23}^2-|\tilde{f}_{13}^{\rm {FC}}|r_{13}^2) \nonumber \\
 &=& 2|\tilde{f}_{13}^{\rm {FC}}|r_{13}^2 g(x,y) , \label{eq:Gdif}
\end{eqnarray}
where
\begin{eqnarray}
g(x,y) &=& \frac{xy \Big\{ \big(\frac{r_{12}}{r_{13}}\big)^2x+\big(\frac{r_{23}}{r_{13}}\big)^2y- 1\Big\}}{x+y-xy}, \\
x &=& -\frac{\tilde{f}_{12}^{\rm {FC}}}{\tilde{f}_{13}^{\rm {FC}}} \ {\rm \ and\ }\  y= -\frac{\tilde{f}_{23}^{\rm {FC}}}{\tilde{f}_{13}^{\rm {FC}}}.
\end{eqnarray}
The force ratios $x$ and $y$ for 
the minimum of $g$ are obtained from $\partial g/\partial x=0$ and $\partial g/\partial y=0$ as
 \begin{equation}
g(x_{\rm {GM}}, y_{\rm {GM}}) = -\frac{r_{12}+r_{23}- \sqrt{(r_{12}+r_{23})^2-r_{13}^2}}{r_{12}+r_{23}+\sqrt{(r_{12}+r_{23})^2-r_{13}^2}}<0 
\end{equation}
with 
 \begin{eqnarray}
x_{\rm {GM}} &=& \frac{r_{12}+r_{23} - \sqrt{(r_{12}+r_{23})^2-r_{13}^2}}{r_{12}} , \\
y_{\rm {GM}} &=& \frac{r_{12}+r_{23} - \sqrt{(r_{12}+r_{23})^2-r_{13}^2}}{r_{23}} . 
\end{eqnarray}
The position ${\bf r}_{\rm m}$ in Eq.~(\ref{eq:rm}) is given by $x_{\rm {GM}}$ and $y_{\rm {GM}}$.
To minimize $\Gamma_{\rm {dif}}$,
the factor $|f_{13}^{\rm {FC}}|$ in Eq.~(\ref{eq:Gdif}) 
is taken as the maximum value while maintaining $|f_{13}^{\rm {FC}}|+|f_{13}^{\rm {CF}}|=|f_{13}^{\rm all}|$, {\it i.e.}, $f_{13}^{\rm {CF}}=0$.
Hence, the lowest value of $\Gamma$ is
obtained for HD with the force center of ${\bf r}_{\rm m}$ and  $f_{13}^{\rm {CF}}=0$.

The local minimum in the region E (LM) is derived from the minimization of $\Gamma + \Gamma_{\rm {CFD}}=-2|\tilde{f}_{13}^{\rm {FC}}|r_{13}^2 g(x,y)$, 
since $f_{12}^{\rm {CF}}f_{12}^{\rm {FC}}<0$, $f_{23}^{\rm {CF}}f_{23}^{\rm {FC}}<0$, and $f_{13}^{\rm {CF}}=0$.
The maximum of $g$ is given at
 \begin{eqnarray}
x_{\rm {LM}} &=& \frac{r_{12}+r_{23} + \sqrt{(r_{12}+r_{23})^2-r_{13}^2}}{r_{12}} , \\
y_{\rm {LM}} &=& \frac{r_{12}+r_{23} + \sqrt{(r_{12}+r_{23})^2-r_{13}^2}}{r_{23}} . 
\end{eqnarray}
Hence, the local-minimum position is determined as ${\bf r}_{\rm lm}= {\bf r}_{\rm 2} - \sqrt{r_{12}r_{23}}\hat{{\bf r}}_{\rm {bv}}$ 
from $x_{\rm {LM}}$ and $y_{\rm {LM}}$.

\section{Simulation Method}

\subsection{Coarse-grained model}\label{sec:simcg}

In the DPD method, the particle motions are integrated in the following Newton's equation
with the DPD thermostat:
\begin{align}
\label{eq:dpd}
 m \dfrac{d{\bf v}_{i}}{dt} &= -\dfrac{\partial U}{\partial {\bf r}_{i}} \\
 &+ \sum_{j \neq i} \left(-w(r_{ij}) {\bf v}_{ij} \cdot \hat{{\bf r}}_{ij} + \sqrt{w(r_{ij})}\xi_{ij}(t)\right) \hat{{\bf r}}_{ij}, \nonumber
 \end{align}
where $U = \sum_{i > j} U_{\mathrm{rep}}(r_{ij}) + \sum_{\mathrm{bonds}} U_{\mathrm{bond}}(r_{ij}) + \sum_{\mathrm{angles}} U_{\mathrm{bend1}}(\theta_{ijk})$
and $w(r_{ij}) = \gamma (1 - r_{ij}/r_{\rm {cut}})$,
with the cutoff at $r_{ij} = r_{\rm {cut}}$  where $\gamma = 4.5 \sqrt{k_{\mathrm B}T m}/r_{\rm {cut}}$.
The Gaussian white noise $\xi_{ij}(t)$ satisfies the fluctuation-dissipation theorem, {\it i.e.}, $\langle \xi_{ij}(t) \rangle = 0$ 
and $\langle \xi_{ij}(t)\xi_{i'j'}(t') \rangle = 2k_{\mathrm B}T(\delta_{ii'}\delta_{jj'} + \delta_{ij'}\delta_{ji'})\delta(t - t')$.

We discretize Eq.~(\ref{eq:dpd}) using Shardlow's S1 splitting algorithm~\cite{shardlow2003splitting}. 
We employ the multi-time-step algorithm
\cite{tuckerman1992reversible,noguchi2007transport}, 
the time step of which, $\Delta t = 0.05\tau$, is different 
from the integration time step $\delta t = 0.005\tau$ for a conservative force $-\partial U/\partial {\bf r}_{i}$,
where $\tau = r_{\rm {cut}} \sqrt{m/k_{\mathrm B}T}$.

All simulations are carried out under the $NVT$ ensemble at the particle density $N/V = 3/r_{\rm {cut}}^3$
with a periodic boundary condition.
The pressure profiles are
calculated for a tensionless membrane at $N_{\mathrm{amp}}=738$, $N_{\mathrm{w}}=9336$, and the side lengths of the simulation box $L_{x}=L_{y}=L_{z}=16r_{\rm {cut}}$
by using the IKN procedure with the decomposition described in Sec.~\ref{sec:dec},
where $N_{\mathrm{amp}}$ and $N_{\mathrm{w}}$ are the numbers of amphiphilic molecules and water particles, respectively.
Amphiphilic molecules are pre-formed into a flat bilayer to reduce the equilibration time.
After the equilibration time $\tau_{\mathrm{eq}} = 10000\tau$ or $15000\tau$, production runs are carried out during $5000\tau$.
The bending rigidity $\kappa$ of the bilayer membrane is estimated at
$N_{\mathrm{amp}} = 2950$ and $N_{\mathrm{w}} = 86504$ by using the
undulation mode of a nearly planar tensionless membrane~\cite{goetz1999mobility,lindahl2000mesoscopic,harmandaris2006novel},
 $\langle |h(q)|^{2} \rangle = k_{\mathrm B}T/\kappa q^{4}$, with the extrapolation of the cutoff wavelength, $q_{\rm cut}\to 0$~\cite{shiba2011estimation},
where $h(q)$ is the Fourier transformation of bilayer height $h(x, y)$.
Error bars are calculated from five independent runs.

\subsection{Atomistic model}\label{sec:simatom}

MD simulations are carried out in the $NPT$ ensemble using the standard version of
GROMACS 5.1 simulation packages~\cite{abraham2015gromacs,pall2014tackling}.
Bilayer membranes consisting of 400 DOPC molecules surrounded by 20000 water molecules 
are simulated under $T=303.15{}^\circ$C and
$P=1 \mathrm{bar}$. The temperature and pressure are controlled by the
Nos\'e-Hoover and Parrinello-Rahman method, respectively. Newton's equation is integrated using the leap-frog algorithm with MD time step $\delta t =
2$ fs. A bond constraint is applied to the bonds with hydrogens using LINCS
algorithm. Long-range electrostatic interactions are calculated via Particle
Mesh Ewald (PME) method. All initial configurations and input parameters are
generated using CHARMM-GUI Membrane Builder~\cite{jo2008charmm,lee2015charmm}.
The total simulation time is 600 ns, and the first $360$ ns is taken as the
equilibration time.

The obtained MD trajectories are fed into a customized version of
GROMACS-LS~\cite{gromacs_ls} to calculate the local stress profiles. The dihedral
contribution to local stress is calculated using CFD. The electrostatic
contribution is calculated using the IKN procedure
with cutoff length $r^{\mathrm{el}}_{\mathrm{cut}}=2.2$nm. 
Venegas {\it et al}. examined the electrostatic contributions to the local pressure
profile using the IKN procedure with finite cutoff by changing
$r^{\mathrm{el}}_{\mathrm{cut}}$
and reported that the local stress profile shows little difference at
$r^{\mathrm{el}}_{\mathrm{cut}} > 2.2$nm~\cite{vanegas2014importance}.

\end{appendix}

%\bibliographystyle{apsrev}
%\bibliography{ref}

\end{document}